# LLM-D12: A Dual-Dimensional Scale of Instrumental and Relational Dependencies on Large Language Models

Ala Yankouskaya*

Department of Psychology, Bournemouth University, Poole, UK, ayankouskaya@bournemouth.ac.uk

Areej B. Babiker

College of Science and Engineering, Hamad Bin Khalifa University, Doha, Qatar, arbabiker@hbku.edu.qa

Syeda W. F. Rizvi

College of Science and Engineering, Hamad Bin Khalifa University, Doha, Qatar, srizvi@hbku.edu.qa

Sameha Alshakhsi

College of Science and Engineering, Hamad Bin Khalifa University, Doha, Qatar, salshakhsi@hbku.edu.qa

Magnus Liebherr

Department of Mechatronics, University Duisburg-Essen, Duisburg, Germany, magnus.liebherr@uni-due.de

Raian Ali

College of Science and Engineering, Hamad Bin Khalifa University, Doha, Qatar, raali2@hbku.edu.qa

There is growing interest in understanding how people interact with large language models (LLMs) and whether such models elicit dependency or even addictive behaviour. Validated tools to assess the extent to which individuals may become dependent on LLMs are scarce and primarily build on classic behavioral addiction symptoms, adapted to the context of LLM use. We view this as a conceptual limitation, as the LLM-human relationship is more nuanced and warrants a fresh and distinct perspective. To address this gap, we developed and validated a new 12-item questionnaire to measure LLM dependency, referred to as LLM-D12. The scale was based on the authors' prior theoretical work, with items developed accordingly and responses collected from 526 participants in the UK. Exploratory and confirmatory factor analyses, performed on separate halves of the total sample using a split-sample approach, supported a two-factor structure: Instrumental Dependency (six items) and Relationship Dependency (six items). Instrumental Dependency reflects the extent to which individuals rely on LLMs to support or collaborate in decision-making and cognitive tasks. Relationship Dependency captures the tendency to perceive LLMs as socially meaningful, sentient, or companion-like entities. The two-factor structure demonstrated excellent internal consistency and clear discriminant validity. External validation confirmed both the conceptual foundation and the distinction between the two subscales. The psychometric properties and structure of our LLM-D12 scale were interpreted in light of the emerging view that dependency on LLMs does not necessarily indicate dysfunction but may still reflect reliance levels that could become problematic in certain contexts.

## 1 INTRODUCTION

The use of large language models (LLMs) has grown at an unprecedented rate, becoming increasingly embedded in both professional and personal life. In 2024, the global LLM market was valued at USD 6.5 billion, with forecasts projecting expansion to over USD 140 billion by 2033, reflecting a compound annual growth rate of 40.7% [1]. This growth is not limited to the commercial sector. By 2025, it is estimated that over 750 million software products will incorporate LLMs, automating approximately 50% of digital workflows [2]. The widespread adoption of these systems reflects not only their functionality but their growing influence on how people search for information, make decisions, and complete cognitively demanding tasks [3], [4]. At first glance, this integration appears wholly beneficial: LLMs streamline problem-solving, improve productivity, and support a wide range of users across domains. However, the very features that make them valuable (e.g., efficiency, fluency, and personalised responsiveness) also can create the conditions for psychological dependency.

Reliance on technology is typically defined as functional trust in a system to carry out tasks based on perceived competence and reliability [5]. Dependency, by contrast, refers to a habitual or automatic form of engagement, often involving diminished autonomy or reduced critical effort [6]. This shift can be understood through the concept of cognitive offloading: when individuals routinely delegate reasoning, memory, or judgement to external systems, this behaviour can become embedded in everyday cognition [7]. When applied to LLMs, offloading is not only encouraged but often rewarded through rapid, seemingly authoritative outputs that reinforce repeated use. Heersmink [8] argues that this process may gradually erode users' own cognitive engagement, particularly in writing, problem-solving, and decision-making. Empirical evidence supports these concerns: frequent AI use has been associated with lower levels of critical thinking, especially among younger users who show higher levels of reliance on LLM-generated answers [9], [10].

What distinguishes LLMs from other technologies is not only their cognitive utility but their capacity to simulate emotional and interpersonal presence [11] [12] [13]. Their conversational design, responsiveness, and apparent sensitivity to user intent may lead individuals to attribute human-like traits to these systems. Anthropomorphisation, already well-documented in human-computer interaction, increases the likelihood of emotional bonding and perceived companionship [14]. Recent studies show that users interacting with anthropomorphic AI agents report significantly higher levels of media dependency, particularly, when those interactions are emotionally satisfying [15]. It has to be noted that OpenAI's system card for GPT-4o acknowledges these risks explicitly, cautioning that users may develop emotional attachments to the model's expressive voice features, potentially reducing their engagement with real-world relationships [16]. Longitudinal findings further suggest that regular, emotionally charged interactions with conversational agents can increase perceptions of empathy and deepen feelings of attachment over time [17]. Research also shows that when LLMs respond to emotionally valenced prompts, their outputs shift in line with users' expectations, reinforcing anthropomorphic illusions and creating the impression of sentient responsiveness [18].

Recent theoretical work has highlighted that LLMs may have the potential to be addictive in ways that are meaningfully different from other digital technologies [19]. For instance, in contrast to platforms such as social media or online gaming, which often promote compulsive use through rewards, rapid stimulation, or social comparison, LLMs appear to encourage repeated engagement through mechanisms such as self-specific feedback, emotional reassurance, and instrumental success. These features can create a time- and cost-effective, high-reward interaction that is perceived as helpful and at times even empathetic. Theoretical analysis has further identified experiences of flow, perceived autonomy, and pseudo-social bonding as central to this emerging form of dependency [19]. These psychological processes not typically addressed by existing behavioural addiction frameworks, but increasingly relevant in understanding user interaction with intelligent systems.

Taken together, the emerging body of evidence suggests that LLM dependency is not simply a more intensive form of digital media use. Rather, it reflects a qualitatively different psychological process involving emotional validation, pseudo-social bonding, and behavioural reinforcement. This contrasts traditional behavioural dependencies such as gaming, social media, or smartphone use that are typically driven by novelty, social feedback, or fear of exclusion [20] [21]. LLMs operate through a different set of incentives: instrumental success [22], perceived empathy [13], and parasocial bounding [23]. As a result, dependency on LLMs may not be perceived as problematic by users themselves, despite potential impacts on autonomy, critical thinking, and social connection.

Although growing awareness surrounds the potential risks associated with dependence on LLMs, tools designed to measure this emerging phenomenon remain relatively limited. One of the earliest contributions is the Problematic ChatGPT Use Scale (PCUS) [24]. Drawing on the diagnostic criteria for Internet Gaming Disorder (IGD) outlined in the DSM-5, the PCUS assesses maladaptive patterns of interaction with ChatGPT. It includes 11 items rated on a 4-point Likert scale, addressing domains such as preoccupation, withdrawal, tolerance, loss of control, conflict, and mood modification. More recently, the AI Chatbot Dependence Scale has been introduced as an 8-item unidimensional measure, focusing on information-seeking behaviour, task execution, and overall intensity of use [25]. This tool captures key aspects of functional engagement, including time spent and cognitive offloading, offering a practical lens through which to examine reliance on conversational AI. However, its scope remains focused primarily on observable usage patterns, without extending to the psychological or affective dimensions of dependency. An alternative perspective conceptualises LLM dependency as a dual-factor construct, distinguishing between two dimensions of dependence - functional and existential [26]. According to this study, functional dependence includes aspects such as efficiency, task performance, and decision support, and is underpinned by the Human-Computer Trust Model, which considers perceived competence, benevolence, reciprocity, and risk [5]. In contrast, existential refers to a more profound psychological attachment, encompassing elements of identity, emotional regulation, and perceived companionship. This component draws upon contemporary theories of addiction to capture deeper forms of reliance [26].

While this multidimensional framework offers a broader understanding of LLM dependency, particularly by acknowledging both instrumental and emotional aspects, it is framed within the language of addiction. As such, it invites reflection on whether constructs like existential engagement can or should be interpreted through a clinical lens, or whether they may represent qualitatively distinct psychological experiences. It has to be noted that clinically defined symptoms of addiction, such as those found in substance use or behavioural disorders, may not fully capture the nuances of LLM dependency. For instance, individuals may turn to LLMs for efficiency, emotional support, or cognitive scaffolding, which do not necessarily reflect pathological use. Applying clinical addiction criteria may therefore risk pathologising behaviours that are adaptive, context-dependent, or rooted in emerging forms of human-technology interaction.

The current study aims to contribute to the growing body of research on assessing dependency on large language models (LLMs). Our conceptualisation of LLM dependency is informed by established psychological frameworks that explain how users engage with technology across cognitive, emotional, and motivational dimensions [19]. One foundational concept is self-specificity, which refers to the capacity of personalised, emotionally resonant interactions to engage an individual's sense of self, thereby increasing subjective relevance and intrinsic motivation [27]. This is supported by Self-Determination Theory, which postulates that the fulfilment of basic psychological needs such as competence, autonomy, and relatedness enhances engagement and psychological well-being [28]. Additionally, our model incorporates the notion of flow, a state of optimal experience characterised by deep absorption, challenge-skill balance, and distorted time perception - conditions often facilitated by adaptive, responsive technologies like ChatGPT [29]. Crucially, we also draw on parasocial interaction theory to account for the formation of emotionally meaningful, one-sided bonds with LLMs [30]. These relationships, although not reciprocal, can facilitate emotional attachment, perceived companionship, and social substitution [31] [32]. By integrating these perspectives, the current study seeks to advance measurement approaches that capture not only the instrumental use of LLMs but also the deeper psychological connections that may underpin patterns of habitual or dependent use.

## 2 METHOD

In this section, we describe the theoretical underpinnings of the developed LLM-D12 scale for measuring dependency on LLMs. We then outline the procedure and method in detail, including the participants, scale items, external validation design, and data analysis.

### 2.1 Theoretical underpinnings

*Cognitive and informational dependency on LLMs*. Despite LLMs offering substantial benefits in enhancing productivity and problem-solving, there is an increasing concern about the user's dependence on these AI systems [19]. Regular usage of these systems can reduce an individual's cognitive abilities and information retention capacity and increase the reliance or dependency on these systems for information [33], [34]. This dependency could contribute to individuals accepting the information provided by them without verifying the output [35], [36]. Individuals may tend to over-trust the information provided by LLMs, even when the answers provided by the system can be erroneous [37]. This is known as automation bias, which may result in a reduction in users' ability to critically evaluate the validity of the information provided [38].

This overreliance on LLMs is amplified by cognitive biases where an individual seeks to find mental shortcuts or heuristics, which may lead to uncritical acceptance of the information provided by LLM systems [39]. This dependency on the information provided by the AI system without validation could create misclassification and misinterpretation [40]. As the information produced by these systems may not always be correct, use of this information for research can pose a risk for plagiarism, fabrication, and falsification [40]. Over time, this dependency can significantly impact human autonomy, including the personal agency and self-confidence of a person [41]. Research suggests lower self-confidence could reduce an individual's critical thinking and cognitive effort, which may further increase dependency on these systems, possibly resulting in uncritical acceptance of the information provided to them [42]. This could be because these systems have redefined the role of AI from simple task automation to collaborative reasoning and problem solving, referred to as "co-intelligence" [31]. The ability for these systems to act as a cognitive partner and help the users in decision-making deepens the potential of individuals to be dependent on these systems.

*Productivity, instant gratification and task-based dependency*. Research indicates that productivity is the primary reason for AI adoption at work, with 57% of respondents citing this in a survey assessing the use of LLM to improve productivity and work quality [43]. LLMs can boost productivity by streamlining tasks by providing real-time support and offering creative solutions [19]. Its use can save time and help users in making faster and more informed decisions. The growing emphasis on increased productivity in workplaces as a measure for performance could have motivated individuals to adopt LLMs for help in their tasks [44]. Productivity is enhanced as LLMs can answer questions quickly, generate content, and retrieve information, saving time and helping individuals to make quicker and more informed decisions [45]. However, the quick responses from these AI systems can reinforce compulsive use and increase dependency for task completion. This is known as the Instant Gratification Effect, where individuals tend to expect things to happen quickly, and with the emergence of LLMs, the users experience quick and personalized responses, making their experience seamless and satisfying [46], [47]. As a result, users' tolerance for delays may reduce, resulting in them preferring LLMs over other ways to gather information or solve problems [48]. Additionally, depending on LLMs for task completion might increase an individual's sense of accomplishment and dopamine-driven satisfaction, which may improve their work-reward cycle [49]. This work-reward cycle may contribute to individuals taking larger workloads, which could lead to burnout, emotional exhaustion, and work-life imbalance, which are predictors of behavioral addiction patterns [50]. As LLM systems can help with task automation and increase productivity, individuals may depend on them to maintain high performance [19].

*Emotional and social dimensions of LLM dependency*. LLMs can simulate human-like conversations, offering companionship and support. As a result, individuals may prefer to engage with LLMs instead of real-life human interaction, which could lead to social isolation and reduced interpersonal skills [51], [52]. This could arise because people may feel more comfortable when they engage with LLMs, due the non-judgmental, accepting language used by LLMs, which facilitates trust and encourages self-disclosure, strengthening the user-chatbot relationship [51],[53]. Such one-sided connections, known as parasocial bonds, involve emotional attachment to an AI system that cannot reciprocate [54], [55]. Research shows that individuals form a bond with AI chatbots and perceive them as social entities that are capable of understanding them [53]. Constant engagement with LLM models can reinforce feelings of familiarity and trust, which may contribute to dependency. Moreover, from perspective of Social exchange theory, people may prefer interacting with LLMs due to the perceived rewards, which may include the immediate support, validation, and companionship provided by these systems, which may outweigh the emotional cost of engaging with these systems [54]. In addition, research has discovered that AI authenticity and anthropomorphism are key drivers of the relationship between humans and chatbots [55]. Taken together, although the human-like qualities of LLMs may offer psychological comfort, they also carry the risk of increasing emotional dependency and reducing real-life social engagement.

*Personalisation, self-specificity and immersive engagement*. LLM systems can engage in human-like conversations with users and provide tailored responses, displaying behaviours that mimic human emotions and social cues [56]. Extending beyond basic information mediators, LLMs involve personalization and the ability to adapt to the users' needs, making communication personally relevant and self-specific [19]. Self-specificity refers to providing the systems with the individual's unique sense of self, which makes the experience with these AI systems personally relevant [57]. This concept aligns with the Social Verification Theory, which suggests that users may engage more deeply with systems that align with an individual's cognitive functioning and emotional experiences through social validation, personalized feedback, or meaningful interaction [58]. Research suggests that individuals may respond more positively to AI systems that mirror their emotional cues with responses that are personalized and relevant to the user's sense of self [59]. Personal relevance and self-specificity can be fostered in several ways. Firstly, LLMs help nurture an individual's sense of self-esteem through the use of respectful and non-offensive language when answering user queries [60]. Secondly, LLMs have the ability to remember previous conversations, which makes the user experience a familiarity effect, fostering trust and validated by the AI system [61]. Thirdly, LLMs are competent in various tasks, which could make the engagement feel more rewarding and encourage the users to engage for a longer time [19]. This dynamic supports the concept of flow in psychology, where users become deeply engaged with the interaction with these AI systems [62]. While flow is typically linked with positive engagement, using LLM for longer durations can contribute to compulsive usage behavior [63]. Over time, this immersive engagement can lead to overreliance on LLMs, highlighting the role of self-specific engagement evolving into dependency on these systems, especially when users seek to replicate the rewarding states associated with the flow states [64].

### 2.2 Participants

***Sample size estimation***. To determine the appropriate sample size, we employed a priory sample size calculation for Structural Equation Models [65]. This calculation was based on several parameters, including the number of observed and latent variables in the model, the expected effect size ($\lambda = 0.10$), the desired level of statistical significance ($\alpha = 0.05$), and the target statistical power ($1 - \beta = 0.80$). The minimum required sample size for model structure was 200 participants. To accommodate both Exploratory Factor Analysis (EFA) and Confirmatory Factor Analysis (CFA), this number was doubled, allowing the dataset to be split randomly between the two procedures. In total, data were collected from 646 individuals.

Several inclusion criteria were set for participants' eligibility for the study. First, they must be 18 years or older, currently residing in the UK, and identifying as British in terms of culture and norms. Second, they have to be familiar with LLM. Several questions were asked to determine their familiarity. Participants who reported that they do not use it, were excluded from the next step of the survey.

**Design**

***Item development***. Our LLM-D12 scale was developed to measure user dependency on LLMs. An initial set of 26 items was generated through an iterative process and a detailed conceptualization of symptoms related to user-reliance on LLMs.

First, based on our previous work [19]. we identified target areas of LLM dependence: personal relevance, parasocial bonding, over-reliance for decision-making and productivity boost and task automation. Six items were generated for each of personal relevance and work-related productivity dimensions while seven items were generated for parasocial bonding and over-reliance for decision-making. Each item was formulated to reflect observed or theorized symptoms of dependency on LLM. The generated items were refined through multiple reviews and evaluated for face validity by two of the authors, both experts in the field and authors of the original work in [19]. Redundant items were removed. All items were rated on a 6-point Likert scale ranging from 1 ("Strongly disagree") to 6 ("Strongly agree"), eliminating a neutral midpoint to encourage participants to express either agreement or disagreement with the statements. This approach increases response variability, which can enhance discriminant validity and improve the robustness of statistical analyses.

In addition, the initial version of the questionnaire, the items referred specifically to "ChatGPT," which was replaced with "LLM" to generalize beyond a single platform, account for variations in participants' preferences, and ensure they responded based on the LLM they personally use. More specifically, the questions were about their declared Primary LLM, i.e. the one they use the most. This aligns with recommendations in scale development literature to ensure stability over time, across situations, and across cases, ensuring responses reflect attitudes toward LLMs broadly rather than a specific product [66].

Second, for each of the four theoretical dimensions, we developed several items (26 items in total) (full list of initial items can be found in Table 1):

1) *Personal relevance* (6 items). This dimension focuses on how users experience the LLMs as enhancing their self-perceived value, confidence and sense of validation as well as understanding their needs. For example, the item: "It understands me well and provides empathetic answers" captures the *positive feedback* of validation described by Yankouskaya et al., [19] wherein users internalize the feeling that LLM's knows them and understands them. The authors argue that LLMs have three distinct features that fuel users' sense of personal relevance: (1) they are designed to avoid offensive language and respond seriously to all user queries, (2) they can remember previous conversations, giving users the impression that the system understands them, reinforcing a positive feedback loop of validation, and (3) they act as competent assistants that help with various tasks, supporting users' need for competence. These abilities to adapt to users' preferences stimulate prolonged engagement and make it more difficult for users to disconnect

2) *Parasocial bounding* (7 items). This dimension measures emotional attachment, parasocial interaction and social displacement. It reflects the emergence of one-sided emotional connections between users and LLMs. Users may treat the LLM as a companion, sharing personal information and developing a sense of connection that mimics aspects of human interaction. This behavior may stem from perceiving LLMs as low-cost, low-effort, and controllable sources of interaction which may lead to reliance on LLMs for emotional and social needs. For example, the item: "I share details about my private life with it", reflects the emotional attachment to LLMs, which is encouraged by their personalized, human-like responses that simulate empathy and social presence, often aligning with the user's viewpoint**.**

3) *Productivity boost and task automation* (6 items). Productivity enhancement and task automation through LLM, as suggested by Yankouskaya et al. [19], may foster dependency by positioning LLMs as essential tools for maintaining efficiency gains. Items in this subscale reflect the perceived benefits of using LLMs to automate routine tasks, streamline workflows and offer creative solutions. The immediate responses provided by LLMs may be perceived as rewarding, initiating a work-reward cycle that motivates users to increase their workload to maintain a sense of accomplishment and satisfaction. For example, item: "It significantly improve my work conditions and enhance my job satisfaction" assesses the role of LLMs in achieving job satisfaction. Over time, users may begin delegating increasingly complex responsibilities to LLMs in order to sustain productivity standards, potentially leading to withdrawal-like symptoms such as frustration and anxiety when the tool is unavailable. Items in this subscale also assess aspects such as work-life balance and changes in workload associated with the use of LLMs.

4) *Over-reliance for decision-making* (7 items). This subscale measures individuals' tendency to rely on LLMs for making decisions, often at the expense of their own judgment. While the use of LLMs as heuristic tools may reduce cognitive load, it also risks diminishing users' self-confidence and personal agency. The subscale items, shaped by concepts such as automation bias and withdrawal symptoms when LLMs are unavailable, reflect trust in the objectivity and speed of LLMs in offering data-driven suggestions. For example, the item: "It is my go-to for assistance in decision-making", captures the salience of LLMs in daily choices. By providing personalized, rapid responses that save time and effort, and are free from emotional or cognitive bias, LLMs may reinforce an addictive behavioural cycle and amplify decision-related anxiety and uncertainty in their absence, effects that may be more pronounced than with task-specific AI tools.

***Pilot testing***. Pilot testing (N = 27) was performed to assess item understanding. Some items showed a right-skewed distribution. Accordingly, the wording was further refined to reduce social desirability bias, make the items less confrontational and enhance clarity [67]. For example, the item "I spend more time on LLM than originally planned" was later revised to a more neutral phrasing: "I stick to the amount of time I originally planned to spend on it, without going over". Therefore, the wording was refined further to reduce social desirability bias and enhance clarity. The final version was rephrased as: "I spend exactly the time and have the conversations I need, without exceeding." This version aimed to balance clarity with neutral framing, minimize potential response bias, and more accurately capture the intended construct.

***Measurements for external validation***. External validation measurements were used to examine the pattern of associations between the dependency subscales (Instrumental and Relationships) and theoretically relevant external variables to assess the convergent validity of the dependency subscales. Internet Addiction (IA), Attitude Toward AI (ATAI) (acceptance and fear), Need for Cognition (NFC) and Trustworthiness of primary LLM were the theoretically relevant validated measures used for the external validation analysis.

*Internet Addiction* was measured using the Internet Addiction Test (IAT-7) including 7 items [60]. TParticipants responded using a 5-point Likert scale where 1 corresponds to never and 5 corresponds to always. The scores for the 7 items were summed to obtain the total score of internet addiction. Therefore, the theoretical range for the total score of internet addiction is 7 to 35. Cronbach's alpha for IA in the present study was 0.826.

*Attitudes towards AI* were measured using the ATAI scale which consists of two subscales: ATAI Acceptance and ATAI Fear [68]. ATAI Acceptance is assessed using 2 items which are measured on an 11-point Likert scale with 0 as strongly disagree and 10 as strongly agree. The total ATAI Acceptance score is calculated by obtaining the sum for the scores of the two items and the theoretical range of the responses was 0 to 20. ATAI fear is assessed using 3 items which are measured on an 11-point Likert scale with 0 as strongly disagree and 10 as strongly agree. The total ATAI Fear score is calculated by obtaining the sum for the scores of the three items and the theoretical range of the responses was 0 to 30. Cronbach's alpha for ATAI in the present study was 0.687 for Acceptance subscale and 0.715 for Fear subscale.

*Participants Need for Cognition* (NFC) was measured using the NFC scale [59]. The responses to the 6 items were measured using a 5-point Likert scale with 1 indicating extremely uncharacteristic of me and 5 indicating extremely characteristic of me. The total score for the need of cognition measured was obtained by summing the score for the 6 items. The theoretical range for NFC score is 6 to 30. Cronbach's alpha was 0.861.

*Trustworthiness of primary LLM*. We also measured perceived Trustworthiness of participants' primary LLM, which was assessed using an 8-item measure, based on a previously established framework [70]. This framework identified eight core dimensions to evaluate the trustworthiness of LLMs (truthfulness, safety, fairness, robustness, privacy, machine ethics, transparency, and accountability). Each dimension was measured using one item. Participants were asked to indicate the extent to which they agree on each dimension and their responses were measured on an 11-point Likert scale ranging from 0 ("not at all') to 10 ("Completely"). For instance, truthfulness of the information provided by the primary LLM was measured using the question "I trust my primary LLM will provide information that is factually accurate and reliable". The scores for the 8 dimensions were summed to obtain the total score for Trustworthiness of primary LLM. The theoretical range for the total score of trust in LLM is 0 to 80. The Cronbach Alpha for this measure was 0.908 which showed excellent reliability for trust in LLM scale.

*Characteristics of LLM usage*. We assessed frequency of LLM use; the method of interaction; the device primarily used for accessing LLM and the typical location of LLM use. The frequency of LLM use was measured on an 11-point Likert scale ranging from 0 ("very unfrequently') to 10 ("very frequently"). The method of interaction with their primary LLM included the following options: text, voice, text and voice, and other. The question about the device they primarily use for accessing their LLM included the following options: PC, mobile, tablet, smart speaker (e.g., Amazon Echo, Gthat particioogle Nest), dedicated AI device (e.g., Humane AI Pin, Rabbit R1), and other. Participants were further asked to report the typical location of LLM use, choosing from: stationary (e.g., at home, office, school), on the move (e.g., while commuting, traveling), both - stationary and on the move, and other.

### 2.3 Procedure

This study was approved by the Ethical Committee at Bournemouth University, UK (N62239, 03.03.2025). The participants were recruited using Prolific online platform (www.prolific.com). The survey was developed and distributed using SurveyMonkey (www.surveymonkey.com). The pilot test was conducted at the end of March 2025. All participants were provided with information about the study and were required to provide informed consent prior to starting the questionnaire. They were also made aware of their right to withdraw from study at any time. Participants who successfully completed the survey received monetary compensation. The main study was conducted between the end of March 2025 and mid-April 2025. The data associated with this work are available at the Open Science Framework (OSF).

### 2.4 Data Analysis

*2.4.1 Data Preprocessing.*

After initial inspection and removing participants who did not complete or failed attention checks, 532 participants (Mean age = 28.17, SD = 5.91, 50.29% males, 48.57% females, 1.14% non-binary and missing the response). Participants who identified themselves as non-binary (2) and missing responses (4) were excluded from the analysis due to small groups size.

*2.4.2 Descriptive statistics.*

The mean, standard deviation, and distribution of participants' responses were calculated for each item to examine central tendency, variability, and overall response patterns.

*2.4.3 Item quality check procedures.*

Prior to entering data into analysis, item quality was assessed by testing the monotonicity assumption [71]. According to this assumption, individuals with higher levels of the underlying trait (e.g., dependency on LLM) should be more likely to select higher response options. Testing this assumption in the present study is essential for two reasons. First, it supports our scale reliability and dimensionality checks because violations of this assumption might indicate that the item doesn't behave consistently across the continuum of all measured items. Second, since items that violate monotonicity may reflect multidimensionality or measurement error, identifying such items at the early stage of analysis helps flag poorly constructed items that may confuse respondents or function differently across groups. The analysis was carried out using the `mokken` package in R [71]. For each item, the analysis generated a scalability coefficient (H), the proportion of violated monotonicity comparisons, the maximum standardised violation (zmax), and a critical flag indicating whether violations exceeded predefined thresholds. Graphical plots of each item were also examined to confirm the consistency of the findings (Supplementary Material 1).

*2.4.4 Exploratory Factor Analysis.*

The aim of the Exploratory Factor Analysis (EFA) was to examine the underlying latent structure of the questionnaire items and to identify coherent factors that reflect the conceptual dimensions of our constructs. The analysis was conducted using R (Version 2024.12.1+563) with the psych and GPArotation packages [72].

Sampling adequacy was assessed using the Kaiser-Meyer-Olkin (KMO) measure. To determine the optimal number of factors, a scree plot and parallel analysis were conducted. Both indicated support for a three-factor solution. As such, a three-factor model was extracted using Principal axis factoring (PAF) with Promax rotation. Principal axis factoring (PAF) was selected as the extraction method due to its suitability for identifying underlying latent constructs when the data may not meet the assumption of multivariate normality. In addition, given the theoretical expectation that psychological constructs are often interrelated rather than orthogonal, Promax rotation provides a more realistic representation of the data structure by allowing for correlations among factors.

Initially, EFA was conducted on the full dataset to gain an understanding of the underlying structure and item behaviour. Items with factor loadings of 0.50 or higher were retained for further analysis, while those falling below this threshold, as well as items that failed to meet the criteria set out in the data quality check procedure, were excluded. Following this initial refinement, a second EFA was performed to re-examine the factor structure based solely on the retained items, with the aim of confirming the stability and interpretability of the emergent factor solution (Supplementary Material 2).

To cross-validate the structure identified in the EFA, we employed network psychometrics, which estimate direct relationships between items rather than latent constructs [73]. Using the R packages qgraph, bootnet, and igraph, we constructed a partial correlation network from the cleaned questionnaire dataset, excluding reverse-coded and poorly performing items. The network was estimated using the EBICglasso algorithm with default hyperparameters ($\gamma = 0.5$) for sparsity and interpretability [74]. Community detection was performed using the walktrap algorithm on an adjacency matrix derived from the weighted network [75]. Centrality metrics (strength, closeness, and betweenness) were extracted to identify influential nodes, and bridge centrality was computed to locate items acting as connectors between communities using the networktools package (Supplementary Material 3).

*2.4.5 Confirmatory factor analysis.*

Following the EFA, a Confirmatory Factor Analysis (CFA) was conducted to test the fit of the hypothesised factor structure derived from the EFA using an independent sample. The CFA was performed using the lavaan package in R (version 0.6-19), employing the maximum likelihood estimation method. This approach was selected due to its capacity to produce reliable parameter estimates under conditions of moderate sample size and approximate normality. The factor structure tested in the CFA was based on the final solution obtained in the second EFA, with only those items retained that met the criteria of loading ≥ .50 and passed all data quality checks. Each latent factor was modelled to load onto its corresponding set of observed variables, and no cross-loadings were permitted, in line with a strict confirmatory approach. Covariances between latent variables were freely estimated to account for potential inter-factor correlations identified during the EFA. Model fit was evaluated using multiple goodness-of-fit indices (absolute, incremental, and parsimonious fit) (Supplementary Material 4).

The chi-square ($\chi^2$) test was used as a measure of exact fit, where a non-significant result indicates a good fit between the model and the observed data. However, given the chi-square test's known sensitivity to sample size, it was interpreted alongside other fit indices. The Comparative Fit Index (CFI) and Tucker-Lewis Index (TLI) were examined as indicators of incremental fit, with values of 0.90 or above considered acceptable, and values of 0.95 or higher indicative of a good model fit. The Root Mean Square Error of Approximation (RMSEA) was also assessed, where values less than or equal to 0.08 suggest an adequate fit and values below 0.06 are indicative of a close fit; the 90% confidence interval around the RMSEA value was also reported. Finally, the Standardised Root Mean Square Residual (SRMR) was considered, with values at or below 0.08 viewed as evidence of acceptable model fit.

The adequacy of the measurement model was further assessed through the evaluation of individual parameter estimates. Standardised factor loadings were examined, with values of 0.50 or higher considered acceptable, and values above 0.70 regarded as indicative of a strong association between observed indicators and their respective latent constructs [76]. The proportion of explained variance ($R^2$) for each item was also reviewed to ensure that observed variables contributed meaningfully to their designated factors. Additionally, the correlations between latent variables were inspected to assess discriminant validity and to ensure that multicollinearity was not present, with correlations below 0.85 considered acceptable [76].

Prior to conducting the confirmatory factor analysis, the data were screened for assumptions relevant to structural modelling. Multivariate outliers were identified using Mahalanobis distance, and cases with p-values below .001 were examined for potential exclusion. The pattern and extent of missing data were examined (1.78%), and data were assumed to be missing at random. Full information maximum likelihood (FIML) estimation was employed to handle missing values, as this approach allows for the inclusion of all available data and yields unbiased estimates under the MAR assumption.

*2.4.6 Reliability Testing.*

To assess the internal consistency of the newly developed scale we conducted a reliability analysis for each subscale (factor) and the total scale score. First, Cronbach's alpha was calculated to estimate the internal consistency of each factor and the full scale. Values above 0.70 were interpreted as acceptable, with higher thresholds (0.80 and 0.90) indicating good and excellent reliability, respectively [77]. Second, we computed item-total correlations for each item to assess how well an individual item correlated with the sum of the remaining items in its subscale. Higher values indicated that the item contributed meaningfully to the construct. Third, we calculated "alpha if item deleted" statistics to determine whether removing any item would improve the overall internal consistency of its respective subscale. This metric helped to identify items that may reduce the reliability of the scale.

*2.4.7 Validation.*

To evaluate the convergent and discriminant validity of the LLM dependency scale, we conducted a set of psychometric analyses using item-level data for each of the two subscales: Instrumental Dependency and Relationship Dependency. Convergent validity was assessed through the calculation of Average Variance Extracted (AVE) and Composite Reliability (CR), while discriminant validity was examined using the Heterotrait–Monotrait (HTMT) ratio (Supplementary Material 5).

External validity of the LLM Dependency subscales was examined using both standardised multiple regression and relative importance analysis (Johnson's weights). Regression models identified variables that uniquely predicted each subscale score, controlling for all other predictors. Relative importance analysis decomposed the total variance ($R^2$) explained by the model, quantifying each predictor's contribution based on both unique and shared effects (i.e., capturing their relative impact).

A multiple linear regression analysis was carried out for each of the subscales to test the external validity of the LLM Dependency Scale. The purpose of these analyses was to examine how well a set of established psychological and behavioural concepts predicted scores on the LLM dependency dimensions. Each regression model included the following variables as predictors: Internet Addiction Score (IAS), AI attitudes (ATAI Acceptance and Fear), need for cognition (NFC) and perceived Trustworthiness of primary LLM. For each model, both unstandardised and standardised coefficients were calculated to allow comparison of relative effect sizes.

To further evaluate the contribution of each external variable in explaining variance in the two dimensions of LLM dependency, we conducted a relative importance analysis using Johnson's weights. Unlike standardised beta coefficients, which may underestimate the contribution of correlated variables, Johnson's relative weights provide a more accurate estimation of the relative impact of predictors by orthogonalising them and calculating their contribution to the outcome variable. Separate analyses were conducted for each of the two subscales of the LLM Dependency Scale.

## 3 RESULTS

### 3.1 Descriptive statistics

An overview of how participants responded to each item in the questionnaire is reported in Table 1. Skewness and kurtosis values were also examined to assess the normality of distribution. Most items displayed acceptable levels of skewness (between -1 and 1) and kurtosis (within ±2), suggesting no severe departures from normality. Item-to-item correlation matrix is reported in Supplementary Material 6.

Table 1: Central tendency and normality of the data (N=526) for the initial 26 items

| Item code | Mean | SD | Skewness | Kurtosis |
|---|---|---|---|---|
| responses_personalized | 4.772 | 0.831 | -0.749 | 1.406 |
| r_confidence_unchanged | 2.817 | 1.189 | 0.451 | -0.331 |
| understands_me | 4.061 | 1.230 | -0.448 | -0.300 |
| less_confident_without | 3.310 | 1.421 | 0.093 | -0.891 |
| immersed_in_it | 3.363 | 1.544 | 0.048 | -1.112 |
| r_spend_exactly_time | 2.738 | 1.263 | 0.593 | -0.261 |
| share_private_life | 2.523 | 1.645 | 0.731 | -0.806 |
| r_no_expressing_emotions | 3.335 | 1.567 | 0.099 | -1.145 |
| genuine_companion | 2.937 | 1.544 | 0.257 | -1.139 |
| feel_less_alone | 2.371 | 1.618 | 0.927 | -0.459 |
| adds_social_life | 2.013 | 1.343 | 1.133 | 0.009 |
| r_solely_tool | 2.401 | 1.562 | 0.900 | -0.420 |

| Item code | Mean | SD | Skewness | Kurtosis |
|---|---|---|---|---|
| no_need_talk_others | 2.059 | 1.395 | 1.147 | 0.162 |
| improves_work | 4.580 | 1.020 | -0.837 | 1.073 |
| more_rewarded | 3.992 | 1.319 | -0.350 | -0.597 |
| empowers_me | 3.829 | 1.431 | -0.319 | -0.878 |
| less_time_life | 2.861 | 1.428 | 0.512 | -0.766 |
| r_same_performance_without | 3.146 | 1.243 | 0.185 | -0.483 |
| more_work_without | 3.943 | 1.400 | -0.474 | -0.605 |
| additional_brain | 4.283 | 1.365 | -0.875 | 0.102 |
| r_less_credible_than_me | 4.002 | 1.127 | -0.419 | -0.307 |
| turn_to_when_able | 3.835 | 1.371 | -0.615 | -0.497 |
| r_same_ease | 3.181 | 1.208 | 0.196 | -0.333 |
| r_equally_certain | 3.027 | 1.201 | 0.246 | -0.494 |
| my_go_to | 3.897 | 1.431 | -0.456 | -0.707 |
| decisions_uneasy_without | 2.766 | 1.426 | 0.495 | -0.639 |

## 3.2 Item quality check

Out of 26 items, the majority demonstrated satisfactory scalability and very few violations (Table 2). Twelve items showed strong performance and can be retained confidently in the final scale. Seven items showed slightly more violations but remained within acceptable psychometric limits. These items were included in the next step of analysis with a detailed analysis on how they perform in factoring. The remaining seven items showed violation of monotonicity assumption. Visual inspection of the monotonicity plots confirmed the test leading to our decision to remove these items.

Table 2: Monotonicity testing results for original items

| Item code | H | Percent Violated | zmax | Interpretation |
|---|---|---|---|---|
| adds_social_life | 0.38 | 0 | 0 | Strong |
| feel_less_alone | 0.39 | 0.01 | 7 | Strong |
| my_go_to | 0.37 | 0.01 | 8 | Strong |
| no_need_talk_others | 0.39 | 0.02 | 6 | Strong |
| understands_me | 0.3 | 0.02 | 16 | Strong |
| genuine_companion | 0.36 | 0.02 | 26 | Strong |
| less_confident_without | 0.33 | 0.02 | 15 | Strong |
| turn_to_when_able | 0.33 | 0.03 | 18 | Strong |
| share_private_life | 0.36 | 0.03 | 25 | Strong |
| decisions_uneasy_without | 0.38 | 0.03 | 18 | Strong |
| additional_brain | 0.35 | 0.03 | 17 | Strong |
| empowers_me | 0.3 | 0.03 | 24 | Strong |
| more_rewarded | 0.31 | 0.04 | 54 | Moderate |
| more_work_without | 0.27 | 0.05 | 26 | Moderate |
| r_solely_tool | 0.29 | 0.07 | 41 | Moderate |
| improves_work | 0.27 | 0.08 | 40 | Moderate |
| responses_personalized | 0.23 | 0.08 | 39 | Moderate |
| immersed_in_it | 0.37 | 0.01 | 22 | Moderate |
| less_time_life | 0.24 | 0.03 | 37 | Moderate |
| r_confidence_unchanged | 0.14 | 0.18 | 120 | Problematic |
| r_spend_exactly_time | 0.15 | 0.22 | 143 | Problematic |
| r_same_performance_without | 0.15 | 0.14 | 112 | Problematic |
| r_no_expressing_emotions | 0.18 | 0.14 | 66 | Problematic |
| r_same_ease | 0.18 | 0.25 | 149 | Problematic |
| r_equally_certain | 0.17 | 0.22 | 101 | Problematic |
| r_less_credible_than_me | 0.05 | 0.29 | 147 | Problematic |

Note. H - scalability coefficient, Percent Violated the proportion of violated monotonicity comparisons, the maximum standardised violation (zmax)

### 3.3 Exploratory Factor Analysis

We ran an EFA on a dataset comprising a randomly selected half of the participants (N = 263), using the set of 19 items from the monotonicity check. The parallel analysis suggested 3 factors; however, the third factor was close to eigenvalue=1. Moreover, the 3-factor structure yielded in poor fit (RMSEA =0 .112, 90% CI [0.082, 0.132], TLI = 0. 791, SRMR=0.090, $\chi^2(207) = 712.449$, $p < .001$). The model fit for the two-factor EFA also yielded poor fit (RMSEA = 0.102, 90% CI [0.092 to 0.112], TLI = 0.832, $\chi^2(229) = 498.64$, $p < .001$), suggesting that although the fit slightly improved, the model does not reproduce the observed covariance matrix well.

A detailed inspection revealed that several items either failed to load ≥ .50 on any factor or showed ambiguous patterning (Table 3). These items were removed from the item pool. In addition, several items were removed based on evidence from the residual correlation matrix, which indicates the extent to which the factor model failed to account for observed relationships among items. The following items exhibited notable residual correlations, suggesting local dependence or redundancy: (i) *empowers_me* had a high residual correlation of .26 with *more_work_without*, the largest among all residuals observed in this cluster; (ii) *more_work_without* demonstrated residual correlations of .26 with *empowers_me* and –.11 with *additional_brain*, implying that these items share variance not adequately explained by the current factor solution; (iii) *additional_brain* showed residual correlations of –.11 with *more_work_without* and .15 with *my_go_to*. Although its primary loading was acceptable, the pattern of residuals suggests it may reflect variancassociated with an unmodeled third factor, beyond the two-factor solution currently adopted; (iv) *improves_work* showed residual correlations with *more_rewarded* (.14) also suggests conceptual overlap within a cluster of performance-oriented items.

After removing the items, we reran EFA with 2 and 3 factors. The 3-factor solution still did not show a good fit (RMSEA =0 .100, 90% CI [0.083, 0.118], TLI = 0. 890, SRMR=0.078, $\chi^2(33) = 92.1$, $p < .001$) while 2-factor structure showed acceptable fit (RMSEA =0 .078, 90% CI [0.071, 0.084], TLI = 0. 925, SRMR=0.05, $\chi^2(53) = 73.1$, $p < .001$). Therefore, we decided to retain the 2-factor solution.

Table 3: Loading evaluation in EFA

| Item code | Loading | Communalities (h2) | Concern |
|---|---|---|---|
| *responses_personalized* | *0.4* | *0.22* | *Low loading* |
| *understands_me* | *0.39* | *0.34* | *Low loading and unclear factor alignment* |
| less_confident_without | 0.58 | 0.39 | Moderate communality |
| immersed_in_it | 0.59 | 0.51 | Acceptable |
| share_private_life | 0.8 | 0.64 | Strong loading |
| genuine_companion | 0.72 | 0.59 | Strong loading |
| feel_less_alone | 0.94 | 0.87 | Very strong loading |
| adds_social_life | 0.71 | 0.61 | Strong loading |
| r_solely_tool | 0.67 | 0.42 | Acceptable loading |
| no_need_talk_others | 0.95 | 0.86 | Very strong loading |
| improves_work | 0.7 | 0.42 | Acceptable loading |
| more_rewarded | 0.68 | 0.48 | Acceptable loading |
| empowers_me | 0.74 | 0.51 | Strong loading |
| *less_time_life* | *0.37* | *0.2* | *Low loading and low communality* |
| more_work_without | 0.74 | 0.46 | Strong loading |
| additional_brain | 0.67 | 0.47 | Acceptable loading |
| turn_to_when_able | 0.56 | 0.42 | Moderate loading and communality |
| my_go_to | 0.66 | 0.55 | Acceptable loading |
| decisions_uneasy_without | 0.58 | 0.49 | Acceptable loading |

Note. Problematic items are highlighted in italic

#### *3.3.1 Cross-validation of the EFA using the Network Analysis.*

After removing all problematic items, we performed the Network analysis. We estimated a regularised partial correlation network using the EBICglasso algorithm. This method offers a complementary approach by examining the direct statistical dependencies among observed items rather than positing latent variables. The resulting network was structured, with 18 non-zero edges among 12 nodes, resulting in a moderate density of 0.273.

Community detection using the walktrap algorithm identified two clearly delineated item clusters. The first community consisted of items which aligned closely with the second factor from the EFA and reflect instrumental dependency (Figure 1). The second community grouped items, corresponding to the first EFA factor and representing relationship dependency on LLMs. The overlap between these communities and the EFA factors provides strong structural validation for the two-dimensional latent model.

It has to be noted that centrality analysis revealed that the item *decisions_uneasy_without* had one of the highest strength centrality scores, indicating its importance within the instrumental dependency cluster. Similarly, *feel_less_alone* emerged as a key node within the relationship cluster. Bridge centrality analysis indicated that while the two communities were largely modular, several items served integrative functions across clusters. For example, *adds_social_life* and *more_rewarded* showed elevated bridge strength, suggesting their dual relevance across both domains.

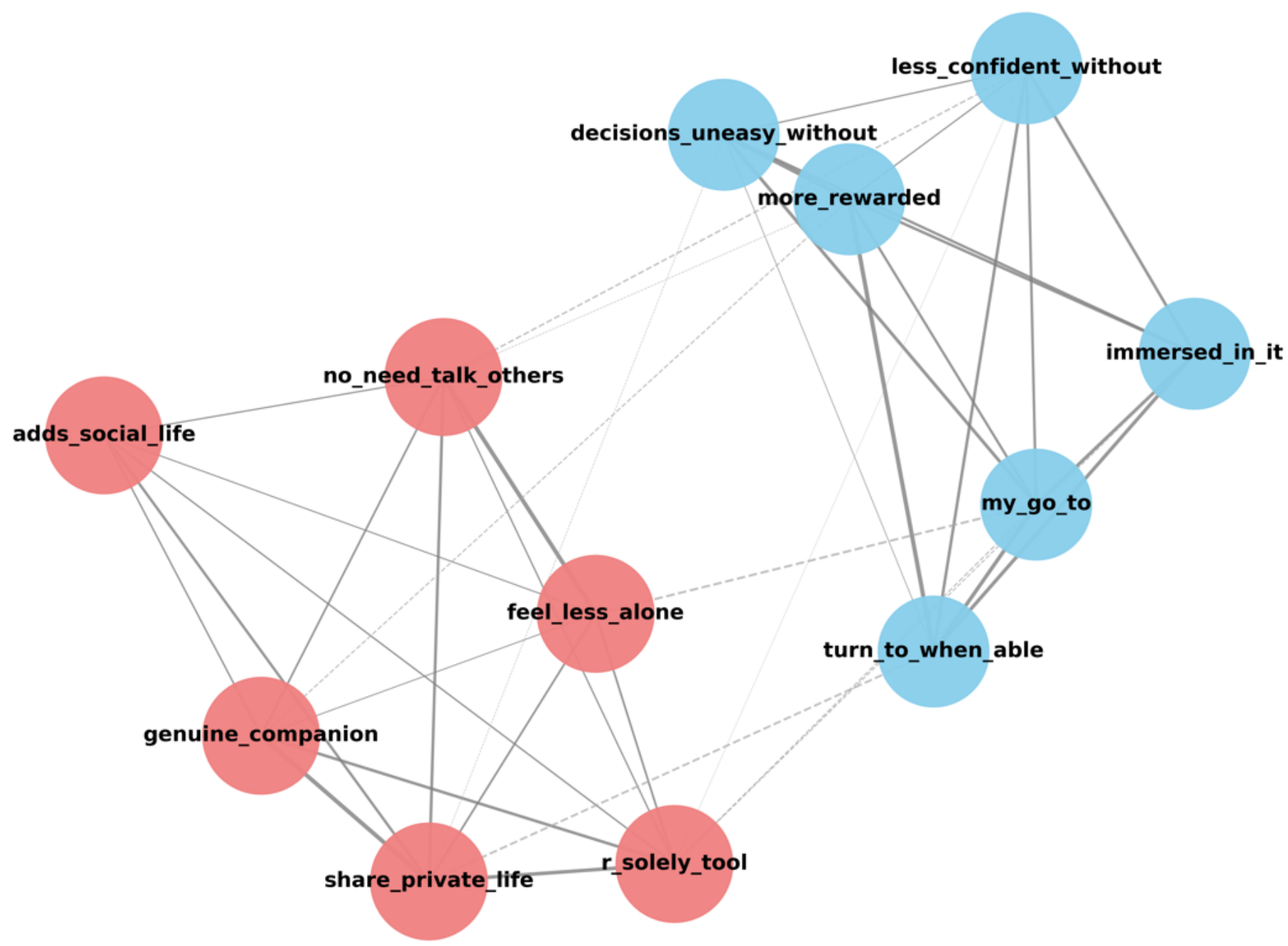

Figure 1: Network Structure of LLM Dependency scale after EFA. The plot displays partial correlation structure among questionnaire items, identifying two distinct item communities: one representing instrumental/task dependency (blue nodes) and the other capturing relationship dependency (red nodes). Solid edges denote stronger within-community associations, while dashed edges indicate inter-community connections

### 3.4 Confirmatory Factor Analysis

The aim of the confirmatory factor analysis (CFA) was to evaluate the factorial structure of our questionnaire using an independent sample (N=263).

The model demonstrated a good overall fit. The chi-square test was significant ($\chi^2 = 133.98$, df = 53, $p < .001$), which is common in large samples or with complex models, and thus not a sole indicator of misfit. The CFI was 0.958 and the TLI was 0.947, both in line with the conventional cutoff of 0.95 for good fit. The RMSEA was 0.076, which is marginally above the commonly accepted threshold of 0.07, but the confidence interval 90% CI [0.060, 0.092] and associated tests suggest it is not a poor fit. The SRMR was 0.046, indicating an excellent fit (values below 0.08 are generally desired). Taken together, these indices suggest that the hypothesised two-factor model provides a reasonably good fit to the data.

All items loaded significantly on their respective latent factors. All items in the first latent factor, labelled here as "Instrumental Dependence", reflect a sense of dependency or reliance on an LLM to support work-related decisions and confidence. Standardised factor loadings ranged from 0.569 to 0.778, all within an acceptable to strong range, and R-squared values showed that each item shared a meaningful proportion of variance with the latent construct (ranging from 32% to 61%) (Table 4). The content of the second latent factor, Relationship Dependency, reflects a more affective, parasocial, and relational dimension in which an LLM as a companion or social presence. Standardised loadings for this factor were also strong, ranging from 0.646 to 0.941, with two items ('feel_less_alone' and 'no_need_talk_others') showing particularly high loadings above 0.9.

Table 4: Factor loadings in the final LLM Dependency Scale

| Item code | Factor | Loading | $R^2$ |
|---|---|---|---|
| less_confident_without | Instrumental Dependency | 0.706 | 0.499 |
| immersed_in_it | Instrumental Dependency | 0.715 | 0.511 |
| more_rewarded | Instrumental Dependency | 0.569 | 0.324 |
| turn_to_when_able | Instrumental Dependency | 0.718 | 0.515 |
| my_go_to | Instrumental Dependency | 0.778 | 0.605 |
| decisions_uneasy_without | Instrumental Dependency | 0.754 | 0.569 |
| share_private_life | Relationship Dependency | 0.791 | 0.625 |
| genuine_companion | Relationship Dependency | 0.756 | 0.572 |
| feel_less_alone | Relationship Dependency | 0.941 | 0.885 |
| adds_social_life | Relationship Dependency | 0.772 | 0.596 |
| r_solely_tool | Relationship Dependency | 0.646 | 0.417 |
| no_need_talk_others | Relationship Dependency | 0.927 | 0.860 |

Table 5 presents the final LLM-D12 Dependency Scale with its two components: Instrumental Dependency (6 items) and Relational Dependency (6 items). The final version of the questionnaire is provided in Appendix 1 and is suitable for use in various contexts, including research. Appendix 2 provides a mapping of individual LLM-D12 items into underlying theoretical constructs and measurements guiding the use of the scale at the item level.

Table 5: LLM-D12: The Large Language Models Dependency 12-item Scale

| **Item code** | **Factor** | **Question** |
|---|---|---|
| **less_confident_without** | Instrumental Dependency | Without it, I feel less confident when making decisions. |
| **immersed_in_it** | Instrumental Dependency | I use it sometimes without realizing how much time I spend immersed in it. |
| **more_rewarded** | Instrumental Dependency | I feel much more rewarded and pleased when completing tasks using it. |
| **turn_to_when_able** | Instrumental Dependency | I turn to it for support in decisions, even when I can make them myself with some effort. |
| **my_go_to** | Instrumental Dependency | It is my go-to for assistance in decision-making. |
| **decisions_uneasy_without** | Instrumental Dependency | Making decisions without it feels somewhat uneasy. |
| **share_private_life** | Relationship Dependency | I share details about my private life with it. |
| **genuine_companion** | Relationship Dependency | I interact with it as if it were a genuine companion. |
| **feel_less_alone** | Relationship Dependency | It helps me feel less alone when I need to talk to someone. |
| **adds_social_life** | Relationship Dependency | It adds to my social life, making socializing more engaging and interesting. |
| **r_solely_tool [R]** | Relationship Dependency | I use it solely as a tool, not to express my feelings or expect it to understand me. |
| **no_need_talk_others** | Relationship Dependency | It helps me feel less alone, reducing the need to talk to others. |

Note. [R] indicates that the item shall be reversed when scoring. Items are rated on a 6-point scale: 1 = Strongly Disagree, 2 = Disagree, 3 = Somewhat Disagree, 4 = Somewhat Agree, 5 = Agree, 6 = Strongly Agree.

### 3.5 Validation

*3.5.1Internal consistency.*

Internal consistency for both factors was high. Cronbach's alpha for Relationship Dependency was 0.91 and for Instrumental Dependency was 0.84, indicating excellent reliability for both scales. All items showed substantial item-total correlations, indicating that each contributes meaningfully to the overall construct (Table 6).

Table 6: Item-level internal consistency of LLM-D12

| **Relationship Dependency Subscale** | | | **Instrumental Dependency Subscale** | | |
|---|---|---|---|---|---|
| Item code | Item-Total correlation | Cronbach's Alpha if item deleted | Item code | Item-Total correlation | Cronbach's Alpha if item deleted |
| share_private_life | 0.765 | 0.893 | less_confident_without | 0.658 | 0.811 |
| genuine_companion | 0.749 | 0.894 | immersed_in_it | 0.580 | 0.828 |
| feel_less_alone | 0.865 | 0.877 | more_rewarded | 0.515 | 0.838 |
| adds_social_life | 0.712 | 0.900 | turn_to_when_able | 0.621 | 0.819 |
| r_solely_tool [R] | 0.607 | 0.915 | my_go_to | 0.714 | 0.800 |
| no_need_talk_others | 0.827 | 0.885 | decisions_uneasy_without | 0.656 | 0.811 |

*3.5.2 Discriminant validity.*

To evaluate discriminant validity between the two latent constructs, we used three tests. First, the Fornell-Larcker criterion was applied by comparing the average variance extracted (AVE) for each construct to the squared inter-construct correlation. The AVE for the first factor, labelled Relationship Dependency was 0.659, and for the second factor, Instrumental Dependency was 0.504. The squared correlation between the two latent variables was 0.503. As both AVE values exceeded the squared correlation, this test supports adequate discriminant validity between the two constructs. Second, the Heterotrait-Monotrait Ratio of correlations (HTMT) was also computed with the resulting HTMT value of 0.589. This is well below the commonly used thresholds of 0.85 or 0.90, further supporting discriminant validity between our latent constructs. Finally, a chi-square difference test was performed to assess whether modelling all items under a single latent factor would provide a comparable fit to the hypothesised two-factor model. The comparison revealed that the one-factor model fit significantly worse than the two-factor solution. Specifically, the chi-square value increased from 133.98 in the two-factor model to 507.74 in the one-factor model, with a difference of 373.76 ($df = 1$, $p < .001$). This deterioration in fit provides further evidence that the two constructs are not interchangeable and should be treated as distinct dimensions.

*3.5.3 External validation.*

The descriptive statistics for the external validation variables are presented in Table 7. These variables include categorical variables (location of use, device used, and interaction method), and continuous variables (frequency of LLM use, Trustworthiness of LLM, Internet Addiction, ATAI Acceptance and Fear, and Need for Cognition).

Table 7: Descriptive statistics for the external validation variables

| | Total Sample (n = 526) | |
|---|---|---|
| **Categorical Variables** | **N** | **%** |
| **Primary LLM Interaction Method** | | |
| Text | 456 | 87.19% |
| Text and Voice | 67 | 12.81% |
| **Primary LLM Device of Use** | | |
| Mobile | 115 | 22.12% |
| PC | 405 | 77.89% |
| **Primary LLM Location of Use** | | |
| Stationary (e.g., at home, office, school) | 350 | 66.67% |
| Both – Stationary and On the Move | 175 | 33.33% |
| **Continuous Measures** | **Mean** | **SD** |
| Frequency of LLM Use | 8.49 | 1.38 |
| Trustworthiness of LLM (Total Score) | 57.40 | 13.63 |
| Internet Addiction (Total Score) | 19.27 | 5.85 |
| ATAI Acceptance (Total Score) | 12.29 | 3.63 |
| ATAI Fear (Total Score) | 13.21 | 5.99 |
| Need for Cognition (Total Score) | 20.62 | 5.03 |

Note: The sums for the two categories in the categorical variables do not add up to the total (526) because responses to other categories were nulled.

To examine the pattern of associations between the dependency subscales and external variables, Pearson correlations were computed. Significant positive correlations were expected with theoretically related constructs such as Trustworthiness of LLM and Internet Addiction. Instrumental Dependency demonstrated strong positive correlations with Trustworthiness of LLM, Internet Addiction, and ATAI Acceptance, but weak negative correlation with ATAI Fear and Need for Cognition (Figure 2).

Relationship dependency showed a similar pattern, being positively associated with Trustworthiness of LLMs, Internet Addiction, and ATAI Acceptance, but not with Need for Cognition. The negative correlation with ATAI Fear was small and at borderline of significance (p=.048) (Figure 2).

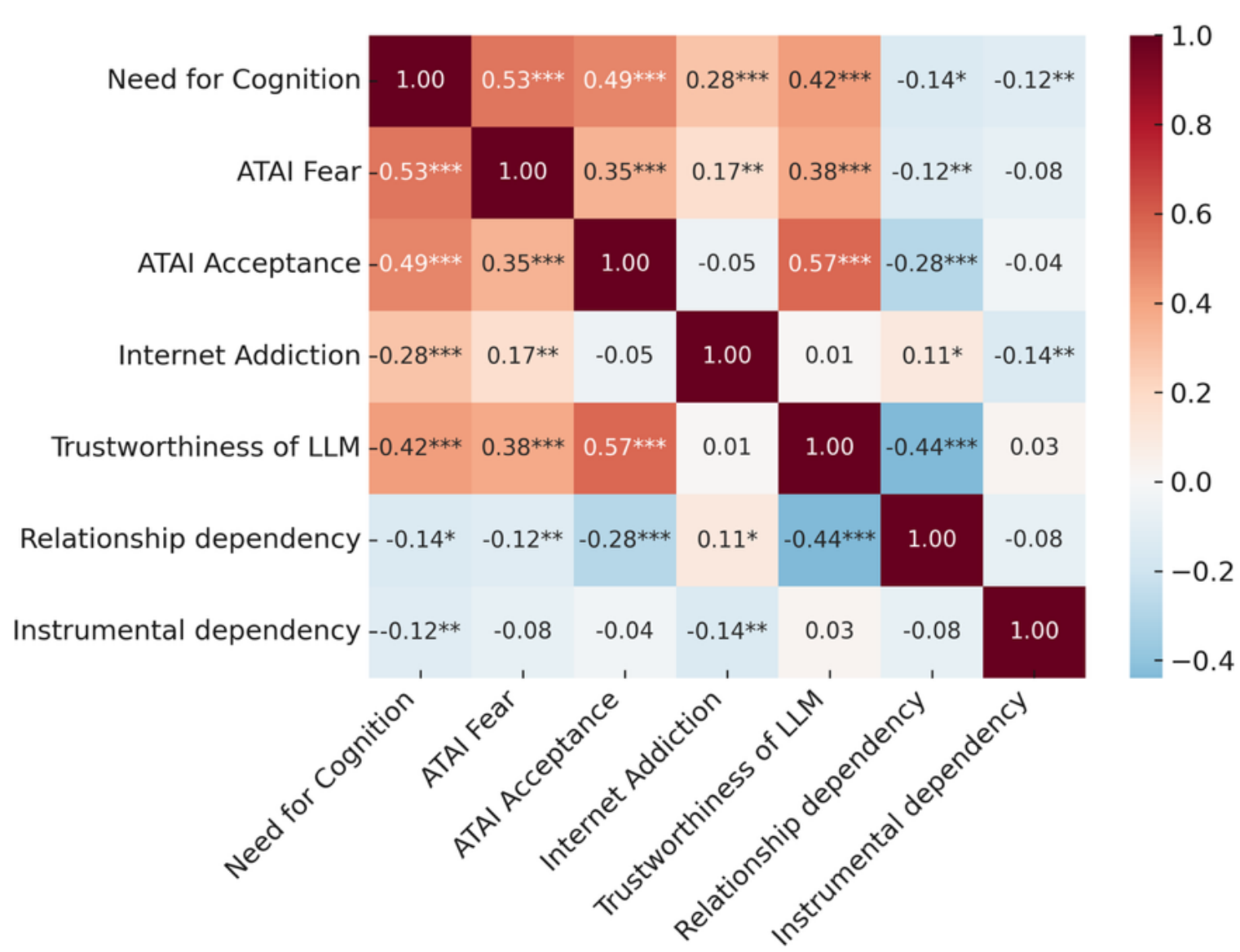

Figure 2: Associations between Instrumental and Relationship Dependency and Trustworthiness of LLM, Internet Addiction, ATAI Acceptance, ATAI Fear and Need for Cognition

We also computed a series of tests to investigate whether the strength of associations between external validation variables differed across the two dependency subscales (Instrumental Dependency and Relationship Dependency). Because both correlations (e.g., Instrumental - Trustworthiness of LLMs and Relationship - Trustworthiness of LLMs) were derived from the same sample of participants, we applied the Steiger's Z-test for dependent (overlapping) correlations [78] implemented in the cocor package in R). The results showed that Instrumental Dependency was more strongly associated with Trustworthiness of LLMs than Relationship Dependency (Steiger's Z = 3.76, p = .0002). Similarly, Instrumental Dependency showed a stronger relationship with internet addiction compared to Relationship Dependency (Steiger's Z = 2.75, p = .006). No difference between Instrumental Dependency and Relationship Dependency were found for ATAI Acceptance Score (Steiger's Z = 1.14, p = .25), ATAI Fear (Steiger's Z = -0.32, p = .75) and Need for Cognition (Steiger's Z = -0.87, p = .39).

We further assessed the validity of our subscales by examining their relationships with theoretically relevant and previously validated constructs (Trustworthiness of LLMs, attitudes toward AI (ATAI: Fear and Acceptance), Need for Cognition, Internet Addiction) as well as key characteristics of LLM usage. Figure 3 provides an overview of the analytical workflow and main findings, followed by a detailed report.

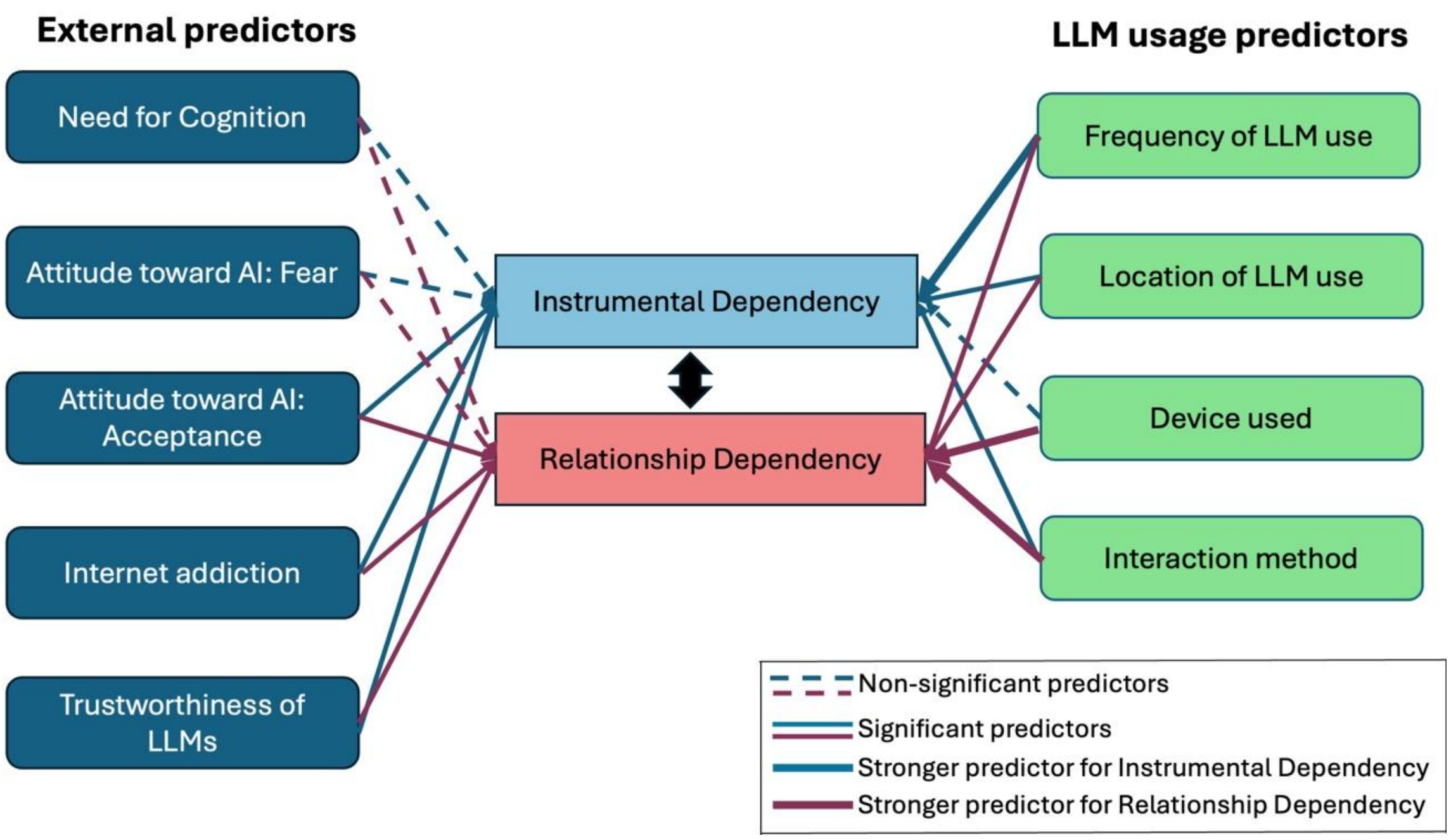

Figure 3: Workflow and main results of external validation

Multivariate regression models were assessed to investigate the unique contributions of each external variable in predicting dependency subscale scores. For the Instrumental Dependency, the regression model was significant, $F(5, 520) = 59.12$, $p < .001$, with an adjusted $R^2$ of .36. Trustworthiness of LLMs ($\beta = 0.18$, $p < .001$), Internet Addiction ($\beta = 0.31$, $p < .001$), and ATAI Acceptance ($\beta = 0.37$, $p < .001$) were significant positive predictors. Need for Cognition showed a trend toward a negative association ($\beta = -0.09$, $p = .054$), while ATAI Fear was not a significant predictor ($\beta = 0.03$, $p = .485$). The regression model for Relationship Dependency explained 20% of variation among the predictors $F(5, 520) = 26.52$, $p < .001$). Trustworthiness of LLMs ($\beta = 0.11$, $p < .001$), Internet Addiction ($\beta = 0.21$, $p < .001$), and ATAI Acceptance ($\beta = 0.57$, $p < .001$) predicted relationship dependency. Neither ATAI Fear ($\beta = 0.04$, $p = .470$) nor Need for Cognition ($\beta = -0.09$, $p = .141$) were significant predictors.

To address the potential threat of common method bias arising from the use of self-report measures, a Harman's single factor test was conducted using confirmatory factor analysis. We tested a model where all indicators loaded onto a single latent factor. The model demonstrated poor fit ($\chi^2(14) = 194.16$, $p < .001$, CFI = .76, TLI = .65, RMSEA = .156, and SRMR = .089) suggesting that no evidence of severe common method bias contaminating the observed relationships. It has to be noted that while not exhaustive, Harman's single factor test widely used as a standard diagnostic tool in self-report questionnaire validation research (e.g., [79] [80]), and our approach aligns with previous studies in applied research (e.g., [81] [82]). Other procedural and statistical remedies, such as the use of a marker variable, the unmeasured latent method factor (ULMF) approach, or temporal and methodological separation, were not feasible in the context of this study. Specifically, temporal separation was not applicable due to the single-session survey format, and the use of a marker variable or latent method factor would have introduced unnecessary model complexity given the early-stage nature of construct development and the absence of established unrelated variables.

In addition, we tested whether LLM usage characteristics predicted scores on the Instrumental and Relationship Dependency subscales, using a multivariate multiple regression. The predictors included: frequency of LLM use, location of use, device used, and interaction method. Prior to the analysis, categories with a small number of responses were excluded. Regarding the location of use, most of the participants responded with 'stationary', followed by 'both – stationary and on the move'. Only one participant responded with 'on the move' and due to the small number of responses for this category, the response was nulled and removed from the subsequent analysis. For device used, many of the participants responded with 'PC', followed by 'mobile' as their device of use for primary LLM. No participants had selected dedicated AI device or smart speaker, and only six participants had responded with 'tablet' as their device of use for the primary LLM. Due to the small number of participants who selected this category, their response was nulled and removed from the subsequent analysis. Lastly, most of the participants reported using 'text' to interact with their primary LLM, whereas a few participants indicated they use 'text and voice' to interact with their primary LLM. Only two participants had selected 'voice', and one participant had selected 'other' interaction method. These responses were nulled and removed from the subsequent analysis. With this removal of categories with small number of responses, we were left with binary options for each of the categorical variables used in the external validation analysis.

A MANOVA using Wilks' Lambda indicated a statistically significant multivariate effect across the two dependency outcomes (Instrumental and Relationship Dependency) for each predictor: frequency of use: $\Lambda = .884$, $F(2, 510) = 33.57$, $p < .001$; location of use: $\Lambda = .896$, $F(2, 510) = 29.59$, $p < .001$; device: $\Lambda = .975$, $F(2, 510) = 6.44$, $p = .002$; interaction method: $\Lambda = .944$, $F(2, 510) = 15.19$, $p < .001$. These results suggest that the set of predictors reliably explained variation in the combined dependency outcomes. Follow-up univariate regressions were conducted for each dependency subscale separately.

The regression model predicting Instrumental Dependency was significant, $F(4, 511) = 24.20$, $p < .001$, accounting for 15.9% of the variance ($R^2 = .159$, adjusted $R^2 = .153$). Frequency of LLM use was a significant positive predictor ($\beta = 1.36$, SE = 0.19, $t = 7.18$, $p < .001$), indicating that more frequent use was associated with higher Instrumental Dependency scores. Individuals who used LLMs stationary reported significantly lower scores of Instrumental Dependency than those who used them in both stationary and mobile settings ($\beta = -1.96$, SE = 0.59, $t = -3.33$, $p < .001$). In addition, users who engaged with LLMs using both text and voice input methods reported higher levels of Instrumental Dependency compared to those who used text only ($\beta = 2.48$, SE = 0.82, $t = 3.03$, $p = .003$). Device type was not a significant predictor in this model ($p = .79$)

The regression model for Relationship Dependency was also significant, $F(4, 511) = 29.97$, $p < .001$, and explained 19.0% of the variance ($R^2 = .190$, adjusted $R^2 = .184$). Frequency of use was a significant positive predictor ($\beta = 0.68$, SE = 0.22, $t = 3.06$, $p = .002$), with individuals who used LLMs more frequently reporting higher Relationship Dependency scores. Those who use LLMs in both stationary and on the move reported significantly higher levels of Relationship Dependency compared to those using it in stationary settings (β = 3.30, SE = 0.69, t = 4.79, $p < .001$). Similarly, individuals using LLMs primarily on mobile devices reported higher Relationship Dependency scores than those using it primarily on PCs (β = 2.21, SE = 0.76, t = 2.93, p = .004). The interaction method was also significant, with users who engaged via both text and voice input reporting substantially higher Relationship Dependency than those using text only (β = 5.23, SE = 0.95, t = 5.49, $p < .001$).

To test whether these predictors had different effects across the two types of dependency, Z-tests were conducted comparing the standardised regression coefficients. Frequency of use was more strongly associated with Instrumental Dependency than with Relationship Dependency (z = 2.35, p = .019). Device type had a significantly greater effect on Relationship Dependency (z = 2.05, p = .040), and interaction method also showed a stronger association with Relationship Dependency (z = 2.19, p = .028). These results support the theoretical distinction between the subscales and suggest that different aspects of LLM interaction are differentially linked to instrumental and relational forms of dependency.

## 4 DISCUSSION

The present study makes an important theoretical and methodological contribution to the field of human–AI interaction by developing and validating a novel scale, LLM-D12, that captures distinct forms of psychological reliance on LLMs. Guided by recent theoretical perspectives on human-AI interaction, the questionnaire was constructed to reflect how LLM use may exhibit symptoms of habitual or preferred reliance [83] [19].Our analysis provides compelling evidence for the psychometric soundness and conceptual clarity of the newly developed LLM Dependency Scale (LLM-D12). This scale captures two distinct but related dimensions of dependency: Instrumental Dependency and Relationship Dependency. The reliability of the LLM-D12 was high across both subscales, demonstrating strong internal consistency, reliability, and discriminant validity.

### 4.1 Instrumental Dependency subscale

The Instrumental Dependency subscale comprises six items that together capture users' reliance on LLMs for cognitive and decision-making support. Each item addresses a distinct but complementary facet of how individuals integrate LLMs into their task management strategies. One important dimension, which may be labeled as dependency on cognitive support, reflects the tendency to feel less confident without the LLM and to experience uneasiness when making decisions in its absence. These behavioural dependencies are well explained by Cognitive Offloading Theory, which proposes that individuals increasingly shift cognitive effort onto external systems to conserve internal resources and reduce uncertainty [7]. A second facet, task absorption through assistance, is measured by the item assessing immersion in the LLM. This form of cognitive engagement where users become absorbed in activities that match their skills and provide immediate feedback, consistent with the principles of Flow Theory [84]. Another dimension, habitual resource seeking, is embodied in turning to the LLM when possible and using it as a habitual 'go-to' resource. This behaviour aligns with Habit Formation Theory, which suggests that repeated successful interactions lead to automatic behavioural patterns over time [85]. All these items showed strong loading (all ≥ 0.70) to the Instrumental Dependency subscale and explained variances were between about 0.50 and 0.61, suggesting that each item shares a substantial proportion of its variance with the factor.

It has to be noted that the item feeling more rewarded through LLM use has a comparatively lower factor loading than the other Instrumental Dependency items. However, its inclusion is theoretically and psychometrically justified. From a conceptual perspective, experiencing reward has been linked to the motivational architecture that sustains cognitive offloading [86], habitual behaviours [87], and decision-making [88]. Because this item taps into a broader motivational process, it may show a slightly lower loading but still plays an important role in explaining why dependency persists over time. For example, it was suggested that such items are often characterised by moderately lower factor loadings because they represent underlying drivers rather than discrete actions [89]. Psychometrically, the item demonstrates acceptable properties. Despite the relatively lower factor loading, it remains well above conventional thresholds for retention, and the squared multiple correlation ($R^2 = 0.324$) indicates a substantial proportion of explained variance. Furthermore, the item showed a strong item-total correlation, and its removal did not result in any improvement in Cronbach's alpha, suggesting that it contributes positively to the internal consistency of the scale.

Based on the content of the Instrumental Dependency subscale, we propose the following definition of this form of LLM dependency as a working construct. Instrumental Dependency is a multidimensional form of overreliance on large language models for cognitive and decision-making support in the context of task management.

### 4.2 Relationship Dependency subscale

The Relationship Dependency subscale captures a novel and increasingly relevant phenomenon: the emergence of parasocial bonds between users and LLMs. The subscale reflects users' cognitive and behavioural tendencies to perceive and engage with LLMs as socially meaningful agents [19]. Each item contributes to this conceptualisation. Specifically, the willingness to share private life with the LLM measures the extent to which users ascribe a degree of confidentiality and responsiveness to the model, a behaviour consistent with parasocial processes identified in media psychology [90]. Another key facet is the attribution of companionship: perceiving an LLM as a genuine companion reflects users' tendency to anthropomorphise and socially personify technology, in line with research showing that individuals form one-sided social bonds with media and AI agents [91]. The subscale also measures the use of LLMs as a means of mitigating social isolation. This reflects users' perceptions that interactions with the LLM help them feel less alone, a phenomenon linked to the social surrogacy hypothesis which proposed that technological interactions can partially substitute for human social contacts [92]. The idea that LLMs add to one's social life further extends this concept, assessing the integration of LLM into an individual's sense of social connectedness [93]. Finally, the reduced perceived need to engage with other people when using LLMs reflects a deeper substitution effect, where LLM interaction is seen as fulfilling some of the roles traditionally occupied by human social partners, a trend also observed in studies of robotic companionship [94].

Psychometric results confirm the coherence of these constructs. All items loaded substantially onto a single relational dependency factor, with especially high loadings observed for perceived reduction in loneliness and social substitution, suggesting that these are central features of parasocial bonding with LLMs. The communalities were consistently high, indicating that the items share a large proportion of their variance with the underlying relational construct. Together, the results support the validity of the Relationship Dependency subscale as a reliable measure of users' emerging parasocial engagement with LLMs.

Based on the scope of the Relationship Dependency subscale, we articulate the following definition of this type of LLM dependency. Relationship Dependency refers to users' cognitive and behavioural tendency to engage with large language models as socially meaningful agents, forming parasocial bonds that reflect perceived companionship, emotional closeness, and social fulfilment.

### 4.3 Instrumental and Relationship Dependency as distinct but related constructs

Our findings indicate that Instrumental and Relationship Dependency represent distinct, though related, dimensions of users' engagement with LLMs. Several lines of evidence support this conclusion. First, the Average Variance Extracted (AVE) for each subscale exceeded the squared correlation between them, suggesting that each construct captures more unique variance than it shares. Second, the HTMT was well below the conventional thresholds for discriminant validity [95], further confirming that the subscales do not simply reflect variations of a single underlying factor. Third, a chi-square difference test demonstrated a significant deterioration in model fit when collapsing the two dimensions into a single latent construct, providing direct empirical support for the superiority of a two-factor solution.

The distinction between Instrumental Dependency and Relationship Dependency reflects well-established ideas in the psychology of human-technology interaction. A growing body of research shows that digital technologies, and increasingly AI systems, are used to meet both practical and social needs, though they do so through different psychological mechanisms and patterns of use [96]. For instance, instrumental engagement typically arises when technologies provide a clear functional benefit (e.g., helping users to make decisions, solve problems, complete tasks more efficiently, generate ideas, or solve complex problems, while also feeling more confident, efficient, or empowered when using it). Such usage aligns with theories of cognitive offloading [97] and distributed cognition [98], suggesting that incorporating these technologies into mental routines can also come with a growing reliance on them [99].

In contrast, Relationship Dependency emerges when users begin to relate to LLM in socially meaningful ways. This form of dependency is not about emotional attachment in the traditional sense, but rather about perceiving the LLM as something more than a tool (e.g., a kind of interactive partner, using LLMs for emotional support, sharing personal experiences with the model, seeking validation or empathy, or consistently attributing human-like qualities). For example, users may turn to the LLM for conversation when feeling isolated, or find comfort in its availability, even when they know it is not truly sentient. Research has shown that people often respond to conversational agents in ways that mirror human interaction, applying social norms such as politeness or reciprocity, and even forming one-sided relationships over time [83] [100].

These parasocial responses are thought to emerge particularly in systems with natural language abilities, where conversational cues encourage users to treat technology as socially present [15]. Relationship Dependency, therefore, reflects a deeper integration of the LLM into the user's social world. However, reliance on LLMs for social connection may, over time, strengthen the dependency on these systems as preferred sources of interaction. This dynamic has been well-documented in the social media literature, where users who turn to digital platforms for social engagement often begin to favour online over offline interactions. For example, one psychological mechanism underpinning this shift is the Compensatory Internet Use model, which suggests that individuals may turn to online technologies to cope with unmet social or emotional needs, and that this coping strategy can gradually displace real-life social engagement [101]. This shift may make them particularly appealing to individuals who experience anxiety, rejection sensitivity, or loneliness in traditional social contexts [102]. In such cases, social reliance on LLMs may move from occasional support to a form of relational dependency, especially if it consistently offers users psychological relief and control.

Taken together, Instrumental Dependency appears to be driven primarily by the perceived cognitive utility of LLMs. While the moderate correlation between the two subscales indicates conceptual relatedness, it also reinforces their distinction, supporting the view that human-LLM engagement encompasses both instrumental and relational forms of interaction. This is consistent with broader perspectives in human-computer interaction, which argue that instrumental and relational uses of technology often coexist and interact [103]. The perception of LLMs as reliable and rewarding instrumental tools may also shift to a gradual attribution of social agency. This is likely because LLM has been designed to mimic human conversational patterns [104]. Similarly, relational engagement may, in turn, reinforce instrumental use by increasing trust and habitual reliance. Thus, the two forms of LLM dependency reflect different but interrelated aspects of users' integration of LLMs into their cognitive and social environments.

### 4.4 External validation of instrumental and relationship dependency

The observed associations between the LLM Dependency subscales and external psychological variables offer support for the theoretical and construct validity of the subscales. The Instrumental Dependency subscale was most strongly associated with trust in LLMs, internet addiction, and positive attitudes towards AI, which aligns with its conceptualisation as a form of cognitive reliance [105]. It has to be noted that trust is central to any instrumental use of algorithmic systems, as users must believe in the system's reliability, accuracy, and utility in order to delegate cognitive tasks [106]. It was identified as a key determinant of technology adoption and integration, particularly when systems are perceived to enhance decision-making or productivity [107]. The association between Instrumental Dependency and trust in AI may explain users' confidence in the tool's competence, reliability, and functional utility [105] [108]. Similarly, the association with internet addiction suggests that habitual or excessive engagement with LLMs may reflect broader digital dependency patterns, particularly when the technology becomes embedded in daily cognitive tasks. However, it is noteworthy that Instrumental Dependency was negatively related to Need for Cognition, albeit weakly. This suggests that individuals with lower intrinsic motivation for effortful thinking may be more likely to offload cognitive tasks to LLMs [109] [7]. Interestingly, our analysis revealed positive but modest association between the Instrumental Dependency and internet addiction. The moderate size of the correlation is in line with our prediction that these constructs share some behavioural overlap but are driven by distinct psychological mechanisms. For instance, Instrumental Dependency appears to reflect users' deliberate integration of LLMs into problem-solving and decision-making routines rather than indication compulsive usage.

The Relationship Dependency displayed a different relationship. While it also correlated with trust and internet addiction, the associations were significantly weaker and unrelated to cognitive effort. This pattern is consistent with previous research suggesting that users can form one-sided social bonds with AI agents in response to unmet social or relational needs [90] [110]. The absence of a significant link with Need for Cognition suggests that Relationship Dependency is not driven by cognitive motivation, but rather by social and emotional context particularly among users who experience LLMs as accessible, low-risk companions.

Importantly, these relationships remained consistent despite tests indicating no evidence of severe common method bias, which enhances confidence that the associations observed are not the result of shared measurement artefacts.

### 4.5 New type of dependency or extension of Internet or social media addiction?

Theoretical and empirical research into behavioural dependency on digital technologies is still in its infancy, that places constraints on how dependency can be defined, measured, and interpreted [111]. Our findings contribute to a growing body of evidence suggesting that LLM dependency may represent a qualitatively distinct form of behavioural reliance, one that is not easily captured by existing models developed for internet addiction or gaming disorder.

While clinical frameworks such as ICD-11 or DSM-5 provide useful starting points and this was recently demonstrated [110], they may overlook critical features specific to LLM use. For example, unlike gaming or general screen-based activity, LLM interactions are typically language-driven, goal-oriented, and cognitively framed, often embedded in contexts such as work, education, or parasocial bounding [19]. The development of our scale suggests that this may be the case. Several items initially included in the LLM Dependency Scale were removed due to poor psychometric performance, including low factor loadings, violations of monotonicity, or high residual correlations. Notably, these excluded items align closely with established symptoms of Internet or Social Media Addiction (e.g., time preoccupation, loss of control, and compensatory use for emotional regulation [97] [112]. For instance, items assessing the perception of spending 'exactly the right amount of time' or expressing uncertainty about performance without the LLM reflect more generalised concerns about digital overuse than LLM-specific behaviours. Similarly, items about being 'empowered' or 'understood' by the LLM can related to a broader gratification or self-related feedback often seen in social media use. Their exclusion from the final factor structure strengthens the argument that LLM Dependency is not simply a subset of existing digital addiction models, but rather reflects a novel behavioural pattern driven by distinct psychological mechanisms such as cognitive offloading, trust in algorithmic assistance, and parasocial engagement with LLM.

It has to be noted that recent studies on AI-based companionship and digital assistants (e.g., Alexa, Replika) show users may form stable, intentional relationships with AI, without necessarily losing control or experiencing harm [113], as required by clinical addiction models. However, what remains unclear is identifying the point at which such companionship shifts from being a supportive interaction to a form of psychological dependency, particularly when the LLM begins to replace rather than complement human relationships. This distinction warrants further conceptual development, particularly in designing diagnostic tools that can differentiate adaptive reliance on LLM from emerging forms of functional or relational overdependence.

Another important angle to consider is whether LLM dependency can be a manifestation of underlying personality traits. Converging models of technology-related behavioural disorders indicate that "addictive-like" engagement often reflects predisposing person factors, most prominently impulsivity and neuroticism, rather than mere exposure to a platform. For example, the I-PACE framework specifies that trait vulnerabilities (e.g., impulsivity, negative affectivity) interact with affective/cognitive responses and executive control to drive compulsive, cue-reactive use [114]. By extension, LLM dependency may arise when these traits meet the immediate, on-demand reinforcement LLMs provide (e.g., rapid feedback, novelty, social-surrogate interaction). However, testing this explicitly is a key next step for construct validation of LLM-dependency scales.

### 4.6 Limitations and future directions

Future research should work towards refining criteria that capture the unique psychological mechanisms and usage patterns associated with LLMs. Our findings indicate that LLM dependency does not follow the typical profile of digital addiction but instead involves distinct instrumental and relational components that reflect users' integration of these systems into everyday thinking and interaction. However, given the cross-sectional nature of the present study, we are unable to determine how these forms of dependency may evolve over time. Longitudinal research will be particularly important in assessing whether reliance on LLMs remains a stable and functional part of users' cognitive and social routines, or whether it develops into a more problematic form of behavioural dependency.

Another limitation of our study is the use of a convenience sample recruited through online platforms. While this approach is widely used in research on digital technologies, especially in early-stage work where participant pools are often self-selecting and interest-driven [115], it restricts the generalisability of the findings [116]. Participants who actively engage with LLMs and voluntarily respond to surveys may not reflect broader user populations, particularly those with limited exposure to or different attitudes towards AI. Moreover, cultural and demographic homogeneity within such samples may obscure important variation in how dependency develops. Although true random sampling in LLM research is difficult to achieve due to the distributed and self-regulated nature of usage, future studies should aim for more diverse and representative samples wherever possible. In particular, testing the structure and expression of Instrumental and Relationship Dependency across different cultural contexts would be an important next step.

Moreover, in the present study we did not collect qualitative data on participants' perception of instrumental and relationship dependency which may have provided additional depth and nuance to the interpretation of LLM dependency. While the focus of this study was on the psychometric development and validation of the proposed subscales, future phases of this work will incorporate qualitative methods to support item refinement and further conceptual clarification.

It has to be noted that the majority of our participants were aged 18–40 years, with only a small proportion being older. This age range was expected to have greater familiarity and engagement with LLMs, as younger and middle-aged adults are more likely to adopt and integrate emerging digital technologies into daily life. However, patterns of use and dependency in adolescents and elderly individuals could differ due to developmental, cognitive, and generational differences in technology adoption. Future research should adapt and validate the present scale for these age groups to ensure its applicability across the lifespan.

A key direction for future research is the cross-cultural validation of the LLM Dependency Scale (LLM-D12). While the current study, conducted with a UK-based sample, provides strong initial evidence for its psychometric validity, this sampling frame may limit the applicability of the findings in a broader global context. Cultural norms strongly influence how people relate to technology, including expectations around autonomy, trust in automation, and attitudes toward socially responsive systems. Such cultural factors can shape both instrumental and relational reliance on LLMs, potentially leading to differences in the expression or structure of LLM dependency across societies. Validating the scale cross-culturally would enable researchers to determine whether LLM dependency reflects a generalisable psychological construct or one that is shaped by culture-specific values, practices, and technology adoption patterns. Moreover, such work could help identify cultural moderators of potentially problematic forms of dependency, contributing to a more nuanced understanding of how human-LLM relationships develop globally. Because our scale is grounded in the core psychological dynamics of human-LLM interaction, it offers a suitable foundation for such comparative work.

Finally, while frequency of use was included as a predictor in our external validation analyses, we acknowledge the broader potential of exploring LLM use patterns, such as time spent, content shared, and types of tasks performed, as a means of generating new insights into user-LLM interaction. We now consider such investigations as a direction for future research focused on understanding behavioural dynamics which can facilitate our understanding of overreliance on LLM.

## 5 CONCLUSION

The LLM Dependency Scale (LLM-D12) is a newly developed and validated tool for assessing users' instrumental and relationship dependency on large language models. By distinguishing between Instrumental Dependency and Relationship Dependency, the scale moves beyond clinically oriented models of internet or social media addiction to capture forms of engagement specific to LLM use. This distinction reflects growing interest in understanding not just whether people overuse AI technologies, but how they come to perceive these systems as cognitively useful or socially meaningful. The scale does not assume that dependency is inherently negative but offers a structure for identifying when patterns of reliance may become problematic. As LLMs become increasingly embedded in personal, academic, and professional settings, LLM-D12 provides a theoretically grounded foundation for future research into the psychological implications of this emerging form of technology engagement. It may also support the development of responsible AI design, user feedback mechanisms, and digital literacy programs aimed at mitigating emerging forms of overdependence.

## ACKNOWLEDGMENTS

Open Access Fund has been provided by Qatar National Library. This publication was supported by NPRP- 14-Cluster Grant # NPRP 14 C- 0916–210015 from the Qatar National Research Fund (a member of the Qatar Foundation). The findings herein reflect the work and are solely the authors' responsibility.

## AUTHORS' CONTRIBUTION

AY, ML, and RA wrote the original paper in [19], which served as the foundation for this work. AB, AY, and RA developed the first version of the scale. RA, AY, and AB revised the scale following the pilot test. ML, SA, and SR reviewed and provided feedback during both phases. SA, AB, RA, and AY curated the data. AY designed and conducted the statistical analysis and reported it along with the discussion. RA contributed to the data analysis. AY, AB, SR and ML wrote the initial draft. SA and RA critically reviewed and revised the manuscript. RA secured the funding.

## SUPPLEMENTARY MATERIAL

The study design, dataset, and analysis files are available on the Open Science Framework (OSF) at the following link: https://osf.io/5fxq7, DOI 10.17605/OSF.IO/5FXQ7

## Appendix 1

### Large Language Model Dependency Questionnaire (LLM-D12)

Thinking specifically about the large language model (LLM) you use most often (such as ChatGPT, DeepSeek, Gemini, Copilot, etc.), please indicate how much you agree or disagree with each of the following statements

| Item nr | | Strongly Disagree | Disagree | Somewhat Disagree | Somewhat Agree | Agree | Strongly Agree |
|---|---|---|---|---|---|---|---|
| ID1 | Without it, I feel less confident when making decisions | | | | | | |
| ID2 | I use it sometimes without realizing how much time I spend immersed in it | | | | | | |
| ID3 | I feel much more rewarded and pleased when completing tasks using it | | | | | | |
| ID4 | I turn to it for support in decisions, even when I can make them myself with some effort | | | | | | |
| ID5 | It is my go-to for assistance in decision-making | | | | | | |
| ID6 | Making decisions without it feels somewhat uneasy | | | | | | |
| RD1 | I share details about my private life with it | | | | | | |
| RD2 | I interact with it as if it were a genuine companion | | | | | | |
| RD3 | It helps me feel less alone when I need to talk to someone | | | | | | |
| RD4 | It adds to my social life, making socializing more engaging and interesting | | | | | | |
| RD-5 [R] | I use it solely as a tool, not to express my feelings or expect it to understand me | | | | | | |
| RD6 | It helps me feel less alone, reducing the need to talk to others | | | | | | |

**Note:**

The LLM examples in the introductory context can be updated to reflect either the popular models of the time or those most relevant to the context in which the scale is administered.

**Scoring:**

Items are rated on a 6-point scale: 1 = Strongly Disagree, 2 = Disagree, 3 = Somewhat Disagree, 4 = Somewhat Agree, 5 = Agree, 6 = Strongly Agree.

[R] indicates that the item shall be reversed when scoring

ID – Instrumental Dependency, RD – Relationship Dependency

Add up the item scores for ID to get the Instrumental Dependency score.

Add up the item scores for RD to get the Relationship Dependency score.

The overall LLM Dependency score is the sum of the two subscale scores.

## Appendix 2

### Mapping individual items of LLM-D12 into underlying theoretical constructs and measurements

| Item ID | Item | Underlying theoretical construct | Measuring the extent of |
|---|---|---|---|
| ID1 | Without it, I feel less confident when making decisions | **Decision-making confidence** | cognitive dependency on the LLM for decision-making confidence |
| ID2 | I use it sometimes without realizing how much time I spend immersed in it | **Immersive engagement** | becoming deeply absorbed in the interaction to the point of reduced awareness of surroundings or schedules |
| ID3 | I feel much more rewarded and pleased when completing tasks using it | **Positive reinforcement from use** | experiencing greater gratification or accomplishment when the LLM is part of the task process |
| ID4 | I turn to it for support in decisions, even when I can make them myself with some effort | **Cognitive offloading** | leaning on the LLM to reduce cognitive effort |
| ID5 | It is my go-to for assistance in decision-making | **Decision-making reliance** | a person turns to the LLM as their primary source of help when making decisions |
| ID6 | Making decisions without it feels somewhat uneasy | **Decision-making unease** | a person experiences discomfort or uncertainty in decision-making when LLM is unavailable (sense of altered decision experience) |
| RD1 | I share details about my private life with it | **Self-disclosure** | a person shares personal or private information with LLM (based on the belief that the interaction is private and non-judgmental) |
| RD2 | I interact with it as if it were a genuine companion | **Companionship** | assigning human-like qualities, intentions, or responsiveness to the LLM in the context of ongoing interaction |
| RD3 | It helps me feel less alone when I need to talk to someone | **Perceived social connectedness** | perceived social connectedness through LLM interaction |
| RD4 | It adds to my social life, making socializing more engaging and interesting | **Positive social enhancement** | belief that one's social life is more engaging, dynamic, or stimulating because of the LLM |
| RD-5 [R] | I use it solely as a tool, not to express my feelings or expect it to understand me | **Withholding personal expression** | one avoids sharing inner states such as opinions, experiences, values, or emotions |
| RD6 | It helps me feel less alone, reducing the need to talk to others | **Social substitution** | LLM interaction replaces interaction with other people because the LLM fulfils that role |

**Note on feasibility of using LLM-D12 at its item level**

Examining responses to individual items in the LLM-D12, rather than relying solely on total subscale scores, can provide a more detailed understanding of the specific behaviours within instrumental and relationship LLM dependency. This approach may be particularly beneficial when:

- Identifying distinct behaviours that could inform targeted interventions.
- Comparing specific aspects of LLM use between different groups or across measurements.
- Determining which item of dependency are most closely associated with other psychological or behavioural variables.
- Conducting exploratory or formative studies where fine-grained behavioural insights are needed.
- Estimating prevalence of individual indicators of dependent behaviour.
- Identifying the hierarchy of indicators of Instrumental or Relationship LLM dependency.

The psychometric properties of the LLM-D12 indicate that the use of individual items as micro-measures is feasible. Both subscales showed good internal consistency and a coherent factor structure, suggesting that each item is meaningfully related to the construct measured by its subscale. However, the following limitations should be considered:

- Each item contributes differently to the corresponding subscale, so item scores cannot be treated as equivalent.
- The relationships between items within each subscale differ in strength. It is recommended to review item loadings and inter-item relationships, such as through factor analysis or network modelling before interpreting individual item scores.
- Interpreting each item as if it measures a separate construct may be misleading, as items within a subscale share variance due to their focus on related aspects of the same underlying concept.